# *Nernst-Ettingshausen effect in thin Pt and W films at low temperatures*

Renjie Luo[1], Tanner J. Legvold[1], Liyang Chen[2], Douglas Natelson[1,3]

[1]Department of Physics and Astronomy, Rice University, Houston TX 77005, USA

[2]Applied Physics Graduate Program, Smalley-Curl Institute, Rice University, Houston TX 77005, USA

[3]Department of Electrical and Computer Engineering and Department of Materials Science and NanoEngineering, Rice University, Houston TX 77005, USA

*Abstract*

As spin caloritronic measurements become increasingly common techniques for characterizing material properties, it is important to quantify potentially confounding effects. We report measurements of the Nernst-Ettingshausen response from room temperature to 5 K in thin film wires of Pt and W, metals commonly used as inverse spin Hall detectors in spin Seebeck characterization. Johnson-Nyquist noise thermometry is used to assess the temperature change of the metals with heater power at low temperatures, and the thermal path is analyzed via finite-element modeling. The Nernst-Ettingshausen response of W is found to be approximately temperature-independent, while the response of Pt increases at low temperatures. These results are discussed in the context of theoretical expectations and the possible role of magnetic impurities in Pt.

While spin caloritronic measurements[1,2] are increasingly common, distinguishing between different physical mechanisms of generating voltages in structures containing magnetic materials and strong spin-orbit coupling metals can be challenging. For example, in the experimental geometry commonly used for local measurements of the spin Seebeck effect (SSE)[3], care must be exercised to distinguish between the SSE and the ordinary Nernst-Ettingshausen response[4,5] of the strong spin-orbit metal used as an inverse spin Hall detector in the SSE measurement. The Nernst-Ettingshausen effect is the generation of a transverse electric field by a longitudinal thermal gradient in the presence of a mutually perpendicular magnetic field, and it has the same dependence on in-plane magnetic field orientation as the SSE in the local geometry. Particularly as spin caloritronic measurements are extended to lower temperatures and more exotic magnetic systems such as candidate spin liquids[6], it is important to have a quantitative sense of the relative magnitude of the Nernst-Ettingshausen response and its temperature dependence.

Here we report measurements of the Nernst-Ettingshausen response in the commonly employed strong spin-orbit metal W and Pt, using a device geometry typically used in local SSE measurements, but in this case fabricated on an inert $SiO_2$/Si substrate, to eliminate any SSE contribution to the measured voltage. We find that the Nernst-Ettingshausen response, linear in magnetic field and heater power, is approximately temperature independent from room temperature to 5 K at fixed heater power for W. In contrast, the response for Pt shows a strong temperature dependence at low temperatures. We discuss a possible explanation for this in the context of prior measurements performed on coinage metals with dilute magnetic impurities. Johnson-Nyquist noise thermometry is used to assess the temperature rise of the metal layers with the application of heater power at low temperatures, and finite element modeling is used to assess the heat flow in the experimental geometry. These results show the need for care when considering local SSE measurements at low temperatures.

Our devices consist of the multilayer structure as shown in Fig. 1a. Photolithography, magnetron sputtering, and liftoff were used to prepare the Pt (W) wire (800 μm long, 10 μm wide, 10 nm thick) on a silicon wafer capped with 300 nm oxide. A lithographically defined $SiO_x$ layer with a thickness of 100 nm was fabricated on the top of the Pt (W) wire by e-beam deposition and liftoff. Finally, a Au wire (1300 μm long, 10 μm wide, 50 nm thick) was deposited on the $SiO_2$ layer, just above the Pt (W) wire, to serve as a heater. The $SiO_x$ layer electrically isolates the Au heater and the Pt (W) wire.

For the Nernst-Ettingshausen response measurements, an AC current at angular frequency ω = 2π × (7.7 Hz) is driven through the Au wire and generates a vertical temperature gradient across the whole device that oscillates at angular frequency 2ω. The voltage across the Pt (W) wire is measured at 2ω using a lock-in amplifier, to detect the voltage response due to the temperature gradient. The measurements are performed as a function of temperature and field in a Quantum Design Physical Property Measurement System (PPMS) equipped with a rotation stage. Example results are shown in Fig. 1. When the device stays at 5 K and the heater power of 1 mW is applied, the second harmonic signal across the W wire has a dominant linear-in-field component that is antisymmetric with the field (Fig. 1b), and this response has the angle dependence expected for the electric field generated by Nernst-Ettingshausen effect, $E_x =$

$\nu B_y \nabla_z T \propto B \cos \alpha$, where $\nu$ is Nernst-Ettingshausen coefficient, $B_y$ is y-component of the field, and $\nabla_z T$ is temperature gradient along z-axis The second harmonic voltage is linear in the heater power up to 2 mW (in Fig. 1d), consistent with a thermal origin. Given the lack of any magnetic response of the SiO$_2$ substrate layer, we identify the observed voltage as originating from the Nernst-Ettingshausen effect in the thin W wire. The similar results can be found in Pt wire, as shown in SM.

The detailed temperature dependence of the magnitude of the Nernst-Ettingshausen response is shown in Fig. 1c. For a temperature sweep at constant heater power, we find that the response of the W wire is almost temperature independent. In contrast, in the Pt wire the response magnitude at constant heater power decreases as temperature is lowered, falling below detectability at temperatures below 150 K, and becoming considerably larger as temperature is further decreased below 20 K.

In this experimental geometry, the Nernst-Ettingshausen signal has previously been considered as approximately temperature independent[7]. The Nernst-Ettingshausen coefficient is $\nu = -\frac{\pi^2}{3} \frac{k_B^2 T}{m^*} \frac{\partial \tau}{\partial \varepsilon}\bigg|_{E_F} = \frac{\pi^2}{3} \frac{k_B^2 T}{e} \frac{\mu}{E_F} \propto T$ for comparatively temperature-independent mobility $\mu$ at low temperatures in thin metal films[4]. At low temperatures, the magnitude of the temperature gradient across the metal layer at a given heater power is inversely proportional to the electronic thermal conductivity $\kappa_{el} \sim T$. Thus, at a fixed heater power, the expected temperature dependence of $V_N \propto \frac{\nu}{\kappa_{el}} \propto \frac{T}{T} \propto 1$, consistent with the data in the W wire.

The observed nontrivial temperature dependence of the Nernst-Ettingshausen response in the Pt wire at fixed heater power is therefore surprising. We consider two candidate mechanisms. One possibility is that the Kapitza thermal resistance[8] at the various metal/dielectric interfaces in the stack, such as between the Pt wire and oxidized silicon substrate, may behave differently with temperature than in the W case, such that the temperature profile at fixed heater power is somehow significantly different between the Pt and W devices. We perform Johnson-Nyquist noise thermometry measurements to test this concern, as described below. Another possibility is that the enhanced Nernst-Ettingshausen response at low temperatures may be due to Kondo scattering by dilute ferromagnetic impurities in the Pt. Prior studies in copper and gold wires containing magnetic impurities have shown a qualitatively similar upturn in Nernst-Ettingshausen response at low temperatures[9]. No obvious Kondo-like response in $R(T)$ of the Pt wire is observed, however (Fig. S2), implying that any magnetic impurity scattering is not contributing noticeably to the longitudinal resistive response.

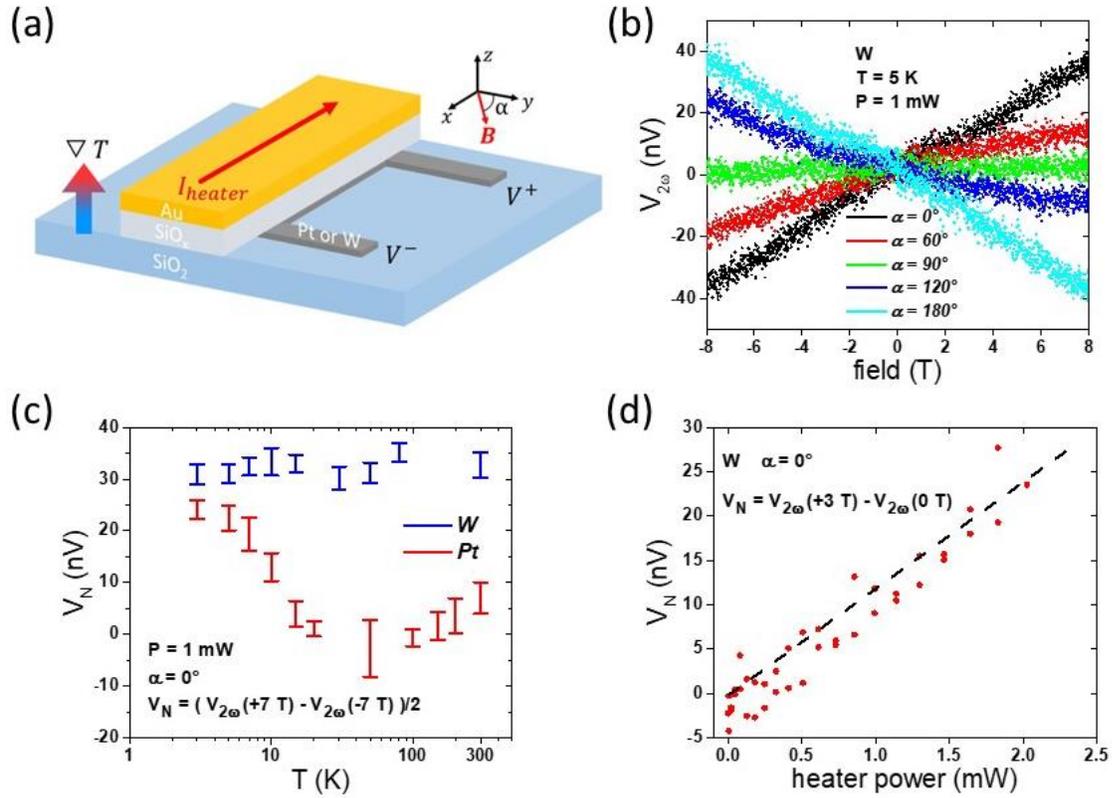

**FIG. 1**. (a) The device geometry in our measurements, showing the definition of field orientation angle $\alpha$ in the *x-y* plane. The heating current runs through the Au wire and generates a temperature gradient perpendicular to the silicon wafer. (b) Field dependence of the second harmonic signal at 5 K in W wire for different field orientation angles. The 2ω signal at 0°/180° gives the largest responses, and the 2ω signal at 90° gives approximately no response, as expected for the Nernst-Ettingshausen effect. (c) Temperature dependence of the Nernst-Ettingshausen voltage in W and Pt at fixed heater power. The Nernst-Ettingshausen voltage is quantified at $\alpha = 0°$ through the difference of the second harmonic signal between 7 T and -7 T, divided by two. The heater power in (b) and (c) is 1 mW. (d) The Nernst-Ettingshausen voltage (quantified as the difference of the second harmonic signal between 3 T and 0 T at $\alpha = 0°$) is linear in heater power, as expected.

To better understand the temperature profile of the device when applying a steady state temperature gradient with the heater, we measure the thermal noise in the Pt (W) wire to obtain the actual temperature change of the detector wire under different heater powers. Johnson-Nyquist thermal noise[10,11], originates from the random motion of charge carriers due to thermal excitation. As a consequence of the fluctuation-dissipation relation, the equilibrium thermal motion of carriers creates a fluctuating voltage on the terminals of an open-circuit resistive element, with a power spectrum density (PSD) given by $S_V = 4k_B TR$, where $T$ and $R$ is the temperature and resistance of the resistive element. As a result, measuring the noise power $S_V$ and the resistance $R$, we can immediately obtain the temperature $T = S_V/4k_B R$. When applying heater power, the temperature of the detector wire will unsurprisingly rise, leading to an increase in $S_V$. Considering that the resistance of the detector wire is approximately temperature-independent below ~ 20 K (Fig. S2), the average temperature increase is $\Delta T = \Delta S_V/4k_B R$. Fig. 2a shows histogram of inferred $\Delta T$ of the Pt wire when the cryostat temperature is fixed at 5 K and a heating power 1 mW is repeatedly applied. Fig. 2b shows that the temperature change $\Delta T$ is slightly sublinear in the heater power, suggesting that at large heater powers some part of the thermal path is driven out of the linear regime of heat transport across the various interfaces. The linearity of the Nernst-Ettingshausen response with heater power (Fig. 1d) implies that the thermal conduction across the metal layer itself remains in the linear regime. Fig. 2c shows a comparison of $\Delta T$ at different cryostat temperatures between the Pt and W wires when applying 1 mW power. The magnitude of $\Delta T$ decreases with the increasing temperature, indicating that thermal conduction of the whole device structure is temperature dependent, as expected when Kapitza-like thermal boundary resistances contribute to the thermal path. The key take-away is that the values of *ΔT* in Pt and W under the same heater conditions are quite similar, implying that the difference in temperature dependence of Nernst-Ettingshausen response (Fig. 1c) is not due to drastically different thermal transport profiles.

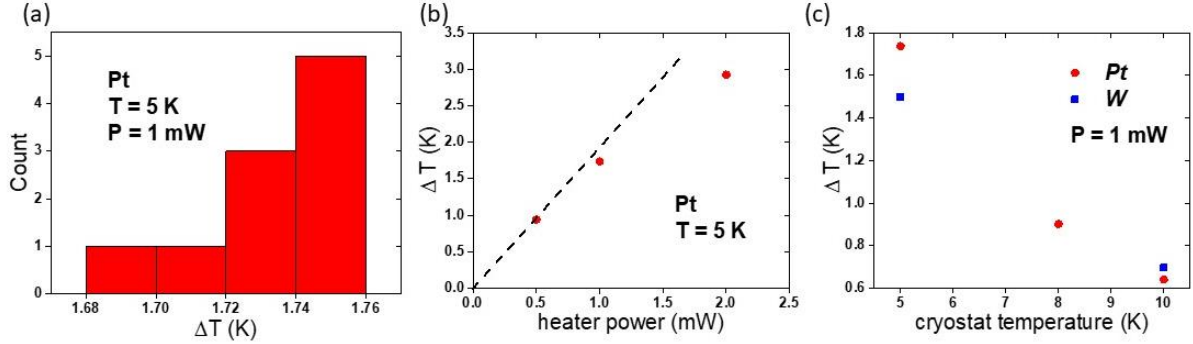

**FIG. 2.** (a) Histogram of the average temperature increase *ΔT* for a Pt wire at cryostat temperature 5 K when a heater power of 1 mW is applied. (b) *ΔT* versus heater power for the Pt device. The black dashed line is a guide to the eye assuming linear response at low heater power. The measured trend is sublinear, suggesting that some component of the complete device thermal path is pushed out of the linear regime of heat transport at high heater powers. (c) *ΔT* at 1 mW heater power versus cryostat temperature, showing that thermal conduction of the whole device changes with temperature, as expected in the presence of Kapitza-like thermal boundary resistance. The error bar in (b) and (c) are smaller than the data point markers.

Nernst-Ettingshausen coefficients in such thin films have not been reported explicitly; however, some control experiments in LSSE measurements[17] allow an estimate of the value in Pt to be $1.5 \times 10^{-11}$ V/(TK) at 10 K (see SM for details). It is worthwhile to do an order-of-magnitude comparison of the measured response with theoretical expectations for the magnitude of the Nernst-Ettingshausen coefficient. Roughly evaluating the simple Fermi liquid expectation[4,5] $\nu = -\frac{\pi^2}{3} \frac{k_B^2 T}{m^*} \frac{\partial \tau}{\partial \varepsilon}\bigg|_{E_F} = \frac{\pi^2}{3} \frac{k_B^2 T}{e} \frac{\mu}{E_F}$ requires values for the Fermi energy $E_F$ and the carrier mobility $\mu$, which may be estimated from the Drude conductivity $\sigma$ and a free electron density $n$ by $\mu = \sigma/(ne)$. In Pt, from the measured resistance and dimensions we find an electrical conductivity at 5 K of $\sigma = 2.31 \times 10^6$ S/m. A reasonable single-band free-electron model estimate for carrier density for Pt[12] is $n \approx 1.6 \times 10^{28}$ m$^{-3}$. This leads to a mobility estimate of $9.02 \times 10^{-4}$ m$^2$/(Vs). From relativistic band structure calculations[13], $E_F \approx 10$ eV. The Nernst-Ettingshausen coefficient for Pt at 5 K should then be $\nu \approx 1.1 \times 10^{-11}$ V/(TK). Similarly, for W, a free electron model estimate for carrier density[14,15] is $n \approx 2.1 \times 10^{28}$ m$^{-3}$, and the Fermi energy[16] is $E_F \approx 9.75$ eV. The measured electrical conductivity of the W film at 5 K is $\sigma = 1.15 \times 10^7$ S/m, giving a mobility estimate of $3.42 \times 10^{-3}$ m$^2$/(Vs). The Nernst-Ettingshausen coefficient for W at 5 K is then estimated to be $\nu \approx 4.3 \times 10^{-11}$ V/(TK).

Extracting the Nernst-Ettingshausen coefficient from the experimental data requires knowledge of the temperature gradient across the metal layer, which we cannot measure directly. A thermal model with reasonable values for material parameters can act as an order-of-magnitude point of comparison with the theory estimates of the Nernst-Ettingshausen coefficient above. Accurate thermal models require the temperature-dependent thermal conductivity of all the involved materials as well as an understanding of interfacial thermal effects. We can estimate the electronic thermal conductivity from the measured electrical conductivity of the metal films using the Wiedemann-Franz law. We take the phonon contribution to be negligible at low temperatures. We neglect the Ti adhesion layer between the Au and the $SiO_x$ isolation layer. We discuss these thermodynamic parameters in detail in the SM.

Much of the total thermal resistance at low temperatures is expected to come from material boundaries[8]. Of particular interest are the metal/dielectric interfaces in the device stack: the boundary between the $SiO_2$ of the substrate and the evaporated Pt (W), the boundary between the Pt (W) and $SiO_x$ isolation layer, and the boundary between the $SiO_x$ and the Au heater. We include these thermal boundary resistances in the model. Unfortunately, even a crude estimate of such interface effects is difficult: phonon boundary resistances depend strongly on the fabrication method, and good comparisons in the literature are sparse[8]. In the model we treat all metal/dielectric interfaces to have the same boundary resistance (valid to within an order of magnitude), which we vary to duplicate the experimentally measured Pt (W) temperature increases at given total heater powers in Fig. 2c. The thermal boundary resistances found at each substrate temperature (see Table S1) are the same order of magnitude as those previously reported[8] and grow with decreasing temperature as expected. From the models we calculate the amount of heat that passes through the Pt (W), and, using the Pt (W) thermal conductivity and thickness, we find the temperature gradient across the metal layer. The model geometry and a representative temperature profile are shown in Fig. 3.

The finite-element model also confirms that a uniform gradient approximation perpendicular to the plane of the device is valid, as the total heat current through the sides of the Pt (W) is roughly 1 µW or less in all cases. Note the Pt (W) wire is only 800 µm long while the Au heater wire is

1300 μm in length, so at a total heater power of 1 mW, only about 0.615 mW of the heater power is directed toward the metal film.

The data in Table S1 and the device geometry are then used to estimate the Nernst coefficient using Eq. (1)

$$\nu \sim \frac{E_x}{B_y}\left(\frac{dT}{dz}\right)^{-1} = \frac{E_x}{B_y}\left(\frac{\kappa_{Pt(W)}\, l\, w}{q_b}\right)^{-1} \quad (1)$$

where $q_b$ is the heat current through the bottom of the metal, $\kappa_{Pt(W)}$ is the Pt (W) thermal conductivity calculated from the measured electrical conductivity, and $l$ and $w$ are the length and width of the Pt (W) wire. For example, the order of magnitude of the Nernst coefficient at substrate temperature 5 K (Pt temperature 6.8K) is $6.9 \times 10^{-11}$ V/(TK). Table S1, S2 show the results for the Pt and W devices in this context.

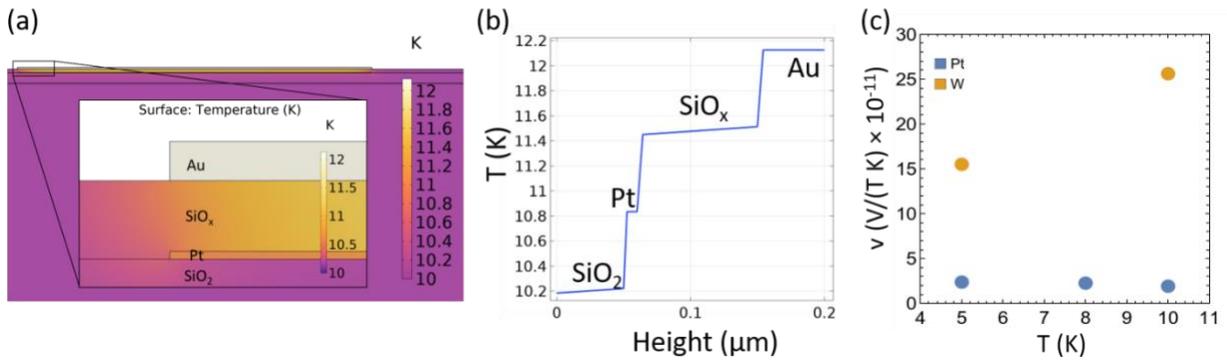

**FIG. 3**. (a) Map of temperature over the cross section of the Pt device at substrate temperature 10 K. (b) example temperature profile vs $z$ within the device at substrate temperature 10 K; the steep interfacial temperature changes result from thermal boundary resistances. (c) Estimated Nernst coefficients for Pt and W at select temperatures.

In summary, we measure the Nernst-Ettingshausen voltage in thin film devices based on Pt and W in the geometry commonly used for local spin Seebeck measurements on magnetic insulators. While the W response is approximately temperature-independent, as expected, the Pt response shows a marked increase at low temperatures. Johnson-Nyquist noise thermometry demonstrates that this difference between the responses of the metals is not a result of some difference in thermal path in the devices, but a property of the metals themselves. We therefore suggest magnetic impurity scattering as the likely explanation for the Pt temperature dependence. A finite element thermal model confirms that both metals show Nernst-Ettingshausen coefficients consistent to an order of magnitude with simple Fermi liquid theoretical estimates. These measurements provide useful bounds for the Nernst-Ettingshausen response as a potential confounding effect in SSE measurements, and the Pt results demonstrate that care must be taken in the assumption that Nernst-Ettingshausen response for metals is constant at low temperatures.

*Acknowledgments*

RL, TJL and DN acknowledge support from DMR-2102028 for the Nernst-Ettingshausen response measurements. LC and DN acknowledge support from DMR-1704264 and DOE BES award DE-FG02-06ER46337 for the noise measurement hardware and software development.

**Author Declarations**

## Conflict of Interest

The authors have no conflicts to disclose.

## Author Contributions

Douglas Natelson: conceptualization (lead), analytical analysis (supporting), funding acquisition (lead), writing (equal); Renjie Luo: investigation (lead), analytical analysis (supporting), writing (equal); Tanner J. Legvold: investigation (supporting), analytical analysis (lead), writing (equal); Liyang Chen: investigation (supporting), writing (supporting).

## Data Availability

The data that support these findings are available on Zenodo (get URL).

Supplementary Materials

for

*Nernst-Ettingshausen effect in thin Pt and W films at low temperatures*

Renjie Luo[1], Tanner J. Legvold[1], Liyang Chen[2], Douglas Natelson[1,3]

[1]*Department of Physics and Astronomy, Rice University, Houston TX 77005, USA*

[2]*Applied Physics Graduate Program, Smalley-Curl Institute, Rice University, Houston TX 77005, USA*

[3]*Department of Electrical and Computer Engineering and Department of Materials Science and NanoEngineering, Rice University, Houston TX 77005, USA*

## S1. Noise measurements

For the noise measurements, we used standard cross correlation methods[1-3] to measure the voltage noise spectra and corresponding voltage noise power spectral density. The programmable voltage source (NI-DAQ 6521) followed by two low-pass LC filters (330 μF and 22 mH for the capacitance and inductance, respectively), providing a clean DC power source for the heater wire. The samples were mounted on a home-built low frequency measurement probe and loaded into the Quantum Design PPMS. Two pairs of twisted wires shielded with stainless steel braided sleeving were used to apply the bias to the heater, collect the noise signal from the Pt (W) wires, and reduce magnetic field induced noise. The sample, LC filters, transmission lines and first pair of pre-amplifiers were shielded by a Faraday cage to reduce environmental noise. The voltage noise across the Pt (W) wire is collected by two identical amplifiers chains, each consisting of two voltage preamplifiers (NF LI-75A, gain = 100, and Stanford Research SR-560, gain = 100). Two amplified voltage signals are recorded by a high-speed oscilloscope (Picoscope 4262) at a sampling rate of 5 MHz. Each spectrum is calculated using cross correlation of two voltage signals containing 400,000 data points, and the spectrum is averaged for 300 times in one measurement.

## S2. Noise measurement calibration

The thermal noise of a resistor at different temperatures were used to calibrate the setup. The voltage noise power spectral density found from the cross-correlation can be expressed as:

$$S_V^m = S_V^a \times A = 4k_B TR \times A$$

where $S_V^a = 4k_B TR$ is the actual Johnson-Nyquist voltage noise power at the resistor $R$, $S_V^m$ is the measured voltage noise power and $A$ is a coefficient containing the squared amplifier gain and a numerical factor related to the cross-correlation parameters of data points of each time series, the sampling frequency and the Hanning window for the Fourier transform.

The thermal noise spectra for a chip resistor of 33.2 Ω at 300K, 150K, and 30K are shown in Fig S1a. The scattered dots are the raw 5 Hz bandwidth spectrum and lines are the averaged spectrum over 1 kHz bandwidth. The spectrum from 100 kHz to 200 kHz is almost flat, so we used the mean value of this region of the spectrum as the measured voltage noise power $S_V^m$. To find the coefficient $A$, we plotted the mean values of $S_V^m$ versus $S_V^a$ in Fig S1b. The noise power at different temperatures fall on the fitting straight line, showing that $S_V^m$ is strictly linearly proportional to $S_V^a$. A linear regression (shown by the red line) finds the relation is:

$$S_V^m = 7.492 * 10^8 \times S_V^a + 1.138 * 10^{-11} \; V^2/Hz$$

The intercept comes from noise background of the measurement system, which brings a constant offset to noise spectrum.

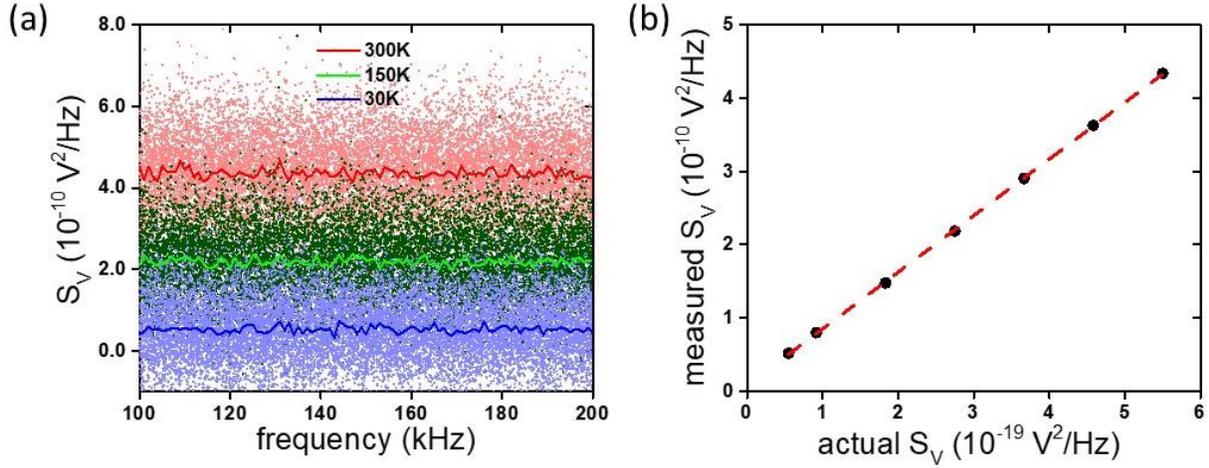

**Figure S1**. Noise measurement setup calibration using the resistor's thermal noise. (a) Thermal noise spectra for the resistor of 33.2 Ω from 100 kHz to 200 kHz, at 3 selected temperatures: 300 K (red), 150 K (green), and 30 K (blue). The scattered dots represent the raw spectra with 5 Hz resolution bandwidth, and solid lines are for the spectra averaged over 1 kHz bandwidth. (b) The measured voltage noise power dependence on the actual voltage noise power given by $4k_BTR$. Linear fitting is used to obtain the effective gain of whole system, and we obtained $7.492 \times 10^8$ for our setup.

## S3. Noise spectrum analysis

The raw noise spectrum rolls off at high frequencies due to capacitance in the wiring and the resistance of the sample. Using standard treatment of capacitive attenuation, the noise at the input end of the first two preamplifiers is:

$$v^2 = v_{na}^2 + \frac{v_s^2}{1 + (R_s C_p \omega)^2}$$

where $v_s$ is the actual voltage noise across the sample, $R_s$ is the sample resistance, $C_p$ is the capacitance, $\omega$ is the frequency, and $v_{na}$ is the input noise from the preamplifiers, which would be eliminated after cross-correlation. This RC model is used to fit the spectrum.

## S4. The resistance of the detector wires at low temperatures.

Fig. S2 shows the temperature dependence of the resistance of the W wire and Pt wire, respectively. Within the range of 5 K to 15 K, the resistance can be considered as approximately independent with the temperature.

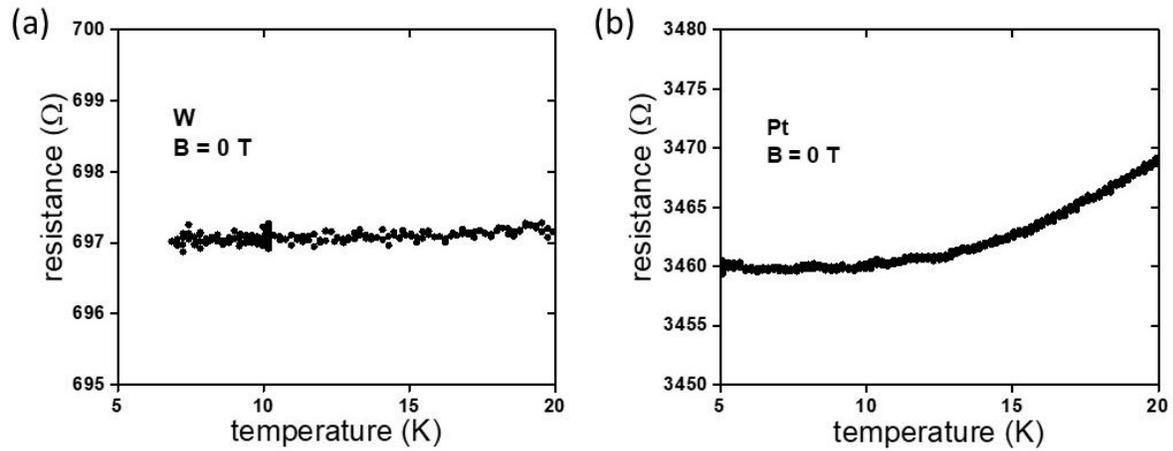

**Figure S2**. Temperature dependence of the resistance of tungsten wire (a) and platinum wire (b), respectively.

## S5. Thermal model parameters and results

The thermal transport model was constructed in COMSOL Multiphysics 6.1 Build 252 with the Heat Transfer Module. The purpose of this model is not to infer rigorous values for the Nernst-Ettingshausen coefficients, as systematic uncertainties in the thermal path are large. The model does provide a consistency check on the magnitude of the measured voltages.

We assume device dimensions are unchanged from those fabricated and measured at room temperature, since thermal contraction effects are all much smaller than other systematic uncertainties in the model.

The thermal conductivity of metals can vary with wire thickness and fabrication techniques, so literatures values are not necessarily comparable. Instead, we used the Wiedemann-Franz law to calculate the electronic contribution from the measured resistivity of the devices.

The thermal conductivity of $SiO_2$ and Si are very different and can depend on thickness, though Ref [4] suggests evaporated $SiO_2$ thickness is beyond size effects. Our Si value comes from Ref [5] while our $SiO_2$ (used for both the evaporated $SiO_x$ layer and the thermal oxide on the Si substrate) value comes from Ref [6].

We insert thermal boundary resistances at all metal/dielectric interfaces and take them to have the same resistance, this should be accurate to within an order of magnitude. In COMSOL this is modeled using a Thermal Contact node with the Equivalent thin resistive layer setting. The nodes resistance is then varied so that the model predicts the same temperature for the Pt (W) as is measured, so the model and real device have the same total thermal resistance to within an order of magnitude. These boundary resistances limit the thermal path, so that the temperature dependence of the bulk $SiO_2$ and Si are comparatively minor influences.

| substrate T (K) | average $T_{Pt}$ (K) | Pt thermal conductivity (W/(mK)) | $R_{metal/dielectric}$ (Kmm²/W) | Heat through top of Pt (mW) | Heat through bottom of Pt (mW) | Estimated temperature gradient (K/m) | Nernst voltage (nV) | Estimated Pt Nernst coefficient, v (V/(TK)) |
|---|---|---|---|---|---|---|---|---|
| 5 | 6.74 | 0.476 | 24 | 0.514 | 0.513 | $135 \times 10^3$ | 22.46 | $2.38 \times 10^{-11}$ |
| 8 | 8.90 | 0.628 | 9.9 | 0.550 | 0.550 | $109 \times 10^3$ | 17.24 | $2.26 \times 10^{-11}$ |
| 10 | 10.6 | 0.751 | 6.1 | 0.578 | 0.578 | $96.2 \times 10^3$ | 12.95 | $1.92 \times 10^{-11}$ |

**Table S1**. Substrate $T$ is set in the experiment, average $T_{Pt}$ is measured, thermal conductivity is calculated, $R_{metal/dielectric}$ tunes the model to give the measured average $T_{Pt}$, heat through top/bottom of Pt is calculated from the model, temperature gradient and Nernst coefficient are then calculated using Eq. 1. The Nernst voltage at 8 K was estimated as a linear interpolation of the 7 K and 10 K Nernst voltages.

| substrate T (K) | average $T_W$ (K) | W thermal conductivity (W/(m K)) | $R_{metal/dielectric}$ (K mm²/W) | Heat through top of W (mW) | Heat through bottom of W (mW) | Inferred temperature gradient (K/m) | Nernst voltage (nV) | Estimated W Nernst coefficient, v (V/(T K)) |
|---|---|---|---|---|---|---|---|---|
| 5 | 6.5 | 2.28 | 20 | 0.521 | 0.521 | $28.6 \times 10^3$ | 31.03 | $15.5 \times 10^{-11}$ |
| 10 | 10.7 | 3.75 | 7.0 | 0.559 | 0.558 | $18.6 \times 10^3$ | 33.36 | $25.6 \times 10^{-11}$ |

**Table S2**. The W equivalent of Table S1.

## S6. Nernst-Ettingshausen estimation from literature results

Ref [7] measured the ordinary Nernst-Ettingshausen response in the longitudinal spin Seebeck effect geometry. Although no specific value of Nernst-Ettingshausen coefficient $\nu$ is reported, we can estimate $\nu$ from the figure. The sample consists of a single-crystalline MgO (5 mm long, 1 mm wide and 1 mm thick) and a Pt thin film (5×1 mm², 7 nm thick). A temperature gradient, $\nabla T$, was generated across the whole device by applying the temperature difference $\Delta T$ between the top of the Pt film and the bottom of MgO. The voltage difference $V$ is measured between the ends of the Pt film while applying an in-plane field, $B$. In accordance with the origin of ordinary Nernst effect and device geometry, $V$ is given by $V = E_N L_x = \nu B \nabla T_{Pt} L_x$. In this configuration, the temperature gradient in the Pt film is different from that in the MgO due to the difference in the thermal conductivities.[8] From the continuous condition at the interface, $\nabla T_{Pt} = (\kappa_{MgO}/\kappa_{Pt})\nabla T_{MgO} \approx (\kappa_{MgO}/\kappa_{Pt})(\Delta T/L_z)$. $\kappa_{Pt}$ is calculated with the Wiedemann-Franz law, using electric conductivity in our measurement, and is 0.75 W/(mK) at 10 K. $\kappa_{MgO}$ is estimated as 100 W/(mK) in the literature[9].

At $T = 10$ K, $\tilde{V}/B$ in the Fig. 1f reads 10 nV/(TK), where $\tilde{V}$ is defined as $V/\Delta T$. Thus, $\nu = (\kappa_{Pt}/\kappa_{MgO})(L_z/L_x)(\tilde{V}/B) = 1.5\times10^{-11}$ V/(TK).

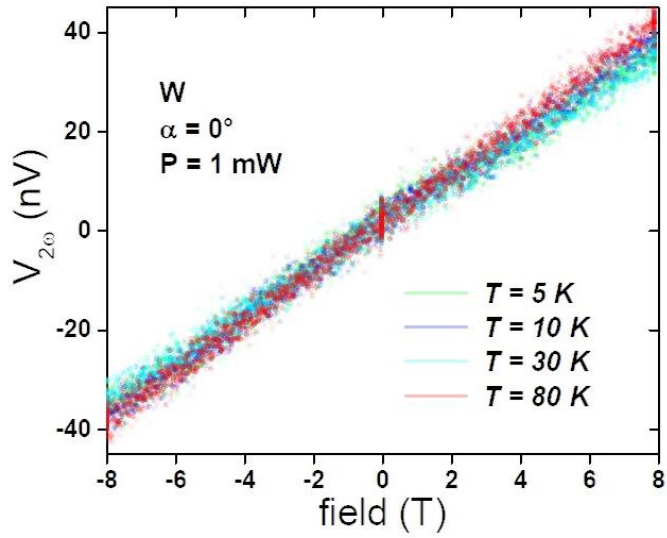

**Figure S3**. Field dependence of the second harmonic signal in W wire at different temperatures. Obviously, the magnitude of the Nernst-Ettingshausen voltage doesn't change a lot.

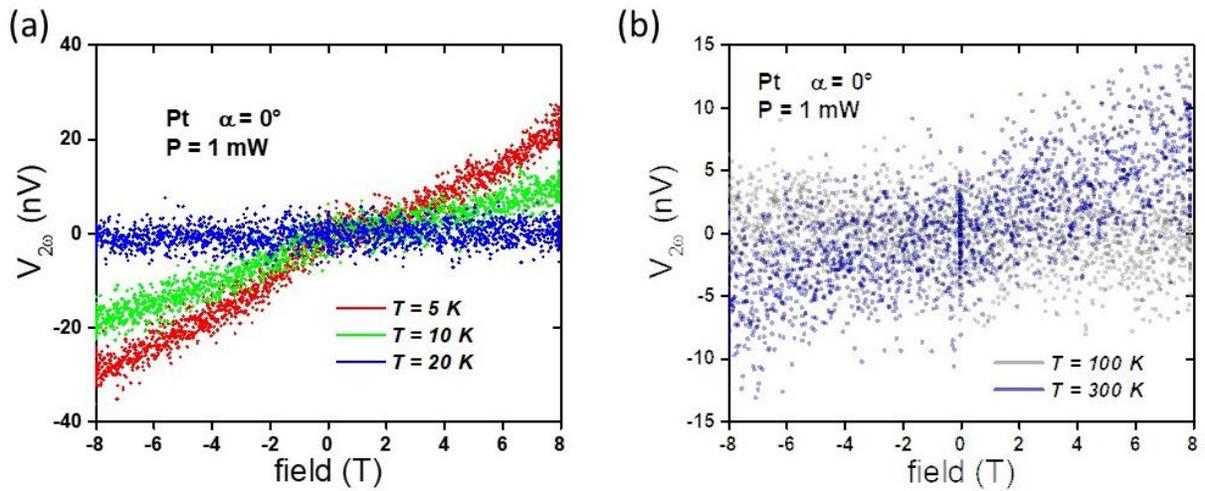

**Figure S4**. Field dependence of the second harmonic signal in Pt wire at different temperatures.

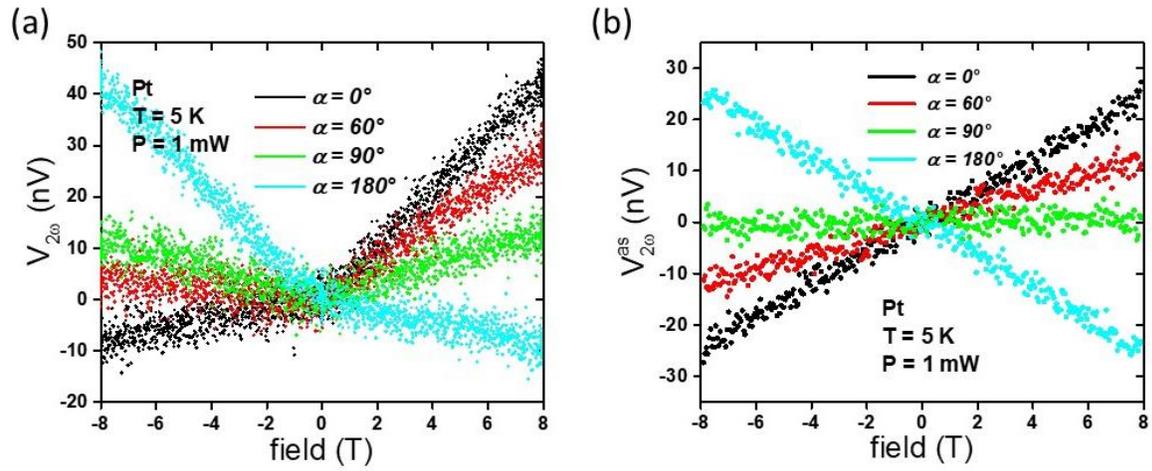

**Figure S5.** (a) Field dependence of the second harmonic signal in Pt wire at 5K with different rotation angles. A much stronger field-symmetric contribution is observed due to parasitic effects in the measurement setup. (b) Antisymmetric component of the second harmonic signal in Pt wire at 5K derived from (a) shows expected angular dependence.